\documentclass[letterpaper, 10 pt, conference]{ieeeconf}  
\IEEEoverridecommandlockouts
\usepackage{cite}
\usepackage{graphicx}
\usepackage{array}
\usepackage{hyperref}
\usepackage{arydshln}
\usepackage{float}
\usepackage{scrextend}
\usepackage{blindtext}
\usepackage{hyphenat}

\let\labelindent\relax

\newcommand{\pfinapp}[2]{%
  \ifx\proofintheappendix\undefined%
	#1%
  \else%
    #2%
  \fi%
}

\usepackage{tikz}
\usepackage{xspace}
\usepackage{xcolor}
\usepackage{amsthm}
\usepackage{amsmath,amssymb,amsfonts}
\usepackage{mathrsfs,mathtools}
\usepackage{mathalfa}
\usepackage{euscript}
\usepackage{csquotes}
\usepackage{enumitem}
\usepackage[normalem]{ulem}

\definecolor{mblue}{rgb}{0,0.4470,0.7410}
\definecolor{morange}{rgb}{0.8500,0.3250,0.0980}
\definecolor{myellow}{rgb}{0.9290,0.6940,0.1250}
\definecolor{mpurple}{rgb}{0.4940,0.1840,0.5560}
\definecolor{mgreen}{rgb}{0.4660,0.6740,0.1880}
\definecolor{mcyan}{rgb}{0.3010,0.7450,0.9330}
\definecolor{mred}{rgb}{0.6350,0.0780,0.1840}
\definecolor{mgreenblue}{rgb}{0.0,1.0,0.5}
\definecolor{parulablue}{rgb}{0.2431,0.1490,0.6588}
\definecolor{parulalblue}{RGB}{39,151,235}
\definecolor{parulagreen}{RGB}{129,204,89}
\definecolor{parulayellow}{RGB}{249,251,21}

\definecolor{cblue}{rgb}{0,0.9,1}
\definecolor{corange}{rgb}{1,0.7,0}

\theoremstyle{definition}
\newtheorem{defn}{Definition}
\newtheorem{exmp}{Example}
\theoremstyle{plain}
\newtheorem{theorem}{Theorem}
\newtheorem{lemma}{Lemma}

\newtheorem{corollary}{Corollary}

\newtheorem{assumption}{Assumption}
\theoremstyle{remark}
\newtheorem{remark}{Remark}

\newenvironment{definition}{\begin{defn}}{\hfill$\square$\end{defn}}

\newcommand{\tss}[1]{\textsuperscript{#1}}

\newcommand{\comment}[1]{}

\newcounter{ass}

\newcommand{\mc}[1]{\mathcal{#1}}
\newcommand{\mf}[1]{\mathfrak{#1}}
\newcommand{\mr}[1]{\mathrm{#1}}
\newcommand{\mb}[1]{\mathbb{#1}}
\newcommand{\ms}[1]{\mathscr{#1}}
\newcommand{\msf}[1]{\mathsf{#1}}
\newcommand{\mt}[1]{\mathtt{#1}}
\newcommand{\mbf}[1]{\mathbf{#1}}
\newcommand{\meu}[1]{\EuScript{#1}}
\DeclareFontFamily{U}{txcal}{\skewchar \font =45}
\DeclareFontShape{U}{txcal}{m}{n}{<-> txr-cal}{}
\DeclareMathAlphabet{\mathcalpxtx}{U}{txcal}{m}{n}

\newcommand{\unaryminus}{\scalebox{0.65}[1]{\ensuremath{\,-}}}

\newcommand{\rankdef}[1]{\ensuremath{\mr{rank}\!\left(#1\right)}}

\newcommand{\rank}{\mr{rank}}
\newcommand{\col}{\mr{col}}
\newcommand{\row}{\mr{row}}
\newcommand{\tdim}{\mr{dim}}

\newcommand{\setdefinition}[2]{\left\{\vphantom{#2}#1\right.\left|\,\vphantom{#1}#2\right\}}

\newcommand{\kron}{\otimes} 

\newcommand{\chris}[1]{\textcolor{mblue}{#1}}

\newcommand{\dnx}{n_\mr{x}}
\newcommand{\dny}{n_\mr{y}}
\newcommand{\dnu}{n_\mr{u}}
\newcommand{\dnp}{n_\mr{p}}
\newcommand{\dnw}{n_\mr{w}}
\newcommand{\dnl}{n_\ell}
\newcommand{\dnr}{n_\mr{r}}
\newcommand{\q}{\mr{q}}
\newcommand{\mdp}[1]{\left(#1 \diamond p\right)}
\newcommand{\KR}{\mf{R}_{\mr{K}}}
\newcommand{\SSR}{\mf{R}_{\mr{SS}}}
\newcommand{\IOR}{\mf{R}_{\mr{IO}}}
\newcommand{\nBf}{\mbf{n}(\mf{B})}
\newcommand{\LBf}{\mbf{L}(\mf{B})}

\newcommand{\Ann}[1]{\mf{N}_{#1}}
\newcommand{\AnnB}[1]{\Ann{\mf{B}}^{#1}}
\newcommand{\Z}{\mb{Z}}
\newcommand{\X}{\mb{X}}
\newcommand{\R}{\mb{R}}
\newcommand{\foralmostall}{\bar{\forall}}
\newcommand{\shiftr}[1]{\overrightarrow{#1}}
\newcommand{\shiftrn}[2]{{\overrightarrow{#1}}^{(#2)}}
\newcommand{\shiftl}[1]{\overleftarrow{#1}}

\newcommand{\Pssobs}[1]{\mc{P}_{\mr{SS},#1}^{(\mr{obs})}}
\newcommand{\Bfint}[2]{\left.\mf{B}_{#1}\right|_{#2}}
\newcommand{\leftpkernel}[2][]{\mr{Kernel}^{\mr{left}}_{\mc{R}, #1 p}(#2)}

\title{Fundamental Lemma for Data-Driven Analysis of\\
                  Linear Parameter-Varying Systems }

\author{Chris Verhoek, Roland T\'oth, Sofie Haesaert and Anne Koch
\thanks{This work has received funding from the European Space Agency (ESA) in the AI4GNC project and from the European Research Council (ERC) under the European Union’s Horizon 2020 research and innovation programme (grant agreement nr. 714663). 
C. Verhoek, R. T\'oth and S. Haesaert are with the Control Systems Group, Eindhoven University of Technology, The Netherlands. R. T\'oth is also with the Institute for Computer Science and Control, Hungary. A. Koch is with the Institute for Systems Theory and Automatic Control, University of Stuttgart, Germany. 
Corresponding author: Chris Verhoek, e-mail: \texttt{c.verhoek@tue.nl}.}%
}

\begin{document}

\maketitle
\thispagestyle{empty}
\pagestyle{empty}

\begin{abstract}
Based on the Fundamental Lemma by Willems et al., the entire behaviour of a Linear Time-Invariant (LTI) system can be characterised by a single data sequence of the system as long the input is persistently exciting. This is an essential result for data-driven analysis and control.
In this work, we aim to generalise this LTI result to Linear Parameter-Varying (LPV) systems. Based on the behavioural framework for LPV systems, we prove that one can obtain a result similar to Willems'. Based on an LPV representation, i.e., embedding, of nonlinear systems, this allows the application of the Fundamental Lemma for systems beyond the linear class.
\end{abstract}

\section{Introduction}\label{sec:introduction}
\noindent Data-driven methods are attractive to obtain system properties or stabilising controllers from data, without identifying a mathematical description of the system itself. One particular result is by Willems et al. \cite{WillemsRapisardaMarkovskyMoor2005}, referred to as the \emph{Fundamental Lemma}, which has been a corner stone for many powerful methods in data-driven analysis and control. This lemma uses the behavioural system theory for (\emph{Discrete-Time} (DT)) \emph{Linear Time-Invariant} (LTI) systems \cite{PoldermanWillems1997} to obtain a characterisation of the system behaviour, based on a single data sequence. More precisely, when one obtains $T$ \emph{input-output} (IO) data points from an LTI system, where the input is \emph{Persistently Exciting} (PE), i.e., the input excited ``all dynamics'' of the system, then the Fundamental Lemma shows that the obtained data spans all possible IO solutions of length $L < T$. 
For LTI systems, this has led to numerous results including (but not limited to) data-based simulation and control \cite{MarkovskyRapisarda2008}, data-driven state-feedback control \cite{markovsky2007linear, dePersisTesi2020}, data-based dissipativity analysis \cite{romer2019one, koch2020provably} and data-driven predictive control \cite{coulson2019data}. There exists preliminary work that aims to extend the Fundamental Lemma towards nonlinear (NL) \cite{alsalti2021data} and Linear Time-Varying (LTV)\cite{nortmann2020data} systems. However, these results impose heavy restrictions on the systems as they leverage model transformations and linearisations. More precisely, these results are modifying the considered system in such a way that on the resulting LTI like description, Willems' Fundamental Lemma can be applied: using feedback linearisation \cite{alsalti2021data}, redefining inputs and outputs for Wiener or Hammerstein systems \cite{BerberichAllgower2020}, using a lumped LTI representation for cyclic LTV systems \cite{nortmann2020data}, or by treating the nonlinearity as a disturbance with a priori known norm bounds \cite{dePersisTesi2020}. Hence, there are no general results for NL systems analogous to those of the Fundamental Lemma.

This paper aims to generalise the Fundamental Lemma to the Linear Parameter-Varying (LPV) system class. LPV systems are linear systems, where the model parameters, describing the linear signal relation, are dependent on a time-varying variable, referred to as the scheduling variable. The latter variable is used to express nonlinearities, time variation, or exogenous effects. The main difference with respect to LTV systems is that the scheduling variable is \emph{not} known a priori; it is only assumed that it is measurable and allowed to vary in a given set. The LPV framework has been shown to be able to capture a relatively large subset of NL systems in terms of LPV surrogate models. Therefore, by extending Willems' result for LPV systems, which is the main contribution of the paper, we make a significant step towards data-driven analysis and control for NL systems.

In \cite{VerhoekAbbasTothHaesaert2021}, some preliminary results on data-driven control for LPV systems with an affine scheduling dependency structure based representation have been introduced using the Fundamental Lemma with additional constraints. In this paper, we obtain results for general LPV systems with representations allowed to have dynamic  \emph{meromorphic} scheduling dependency using the behavioural theory for LPV systems \cite{Toth11_LPVBehav,Toth2010}. These results allow data-driven analysis and simulation for a wide range of LPV representation forms and scheduling dependencies. Moreover, as an additional contribution of the paper, we show that the results in \cite{VerhoekAbbasTothHaesaert2021} are a special case of the developed theory.

The paper is structured as follows. The problem statement in Section \ref{sec:problemstatement} is followed by a presentation of the mathematical building blocks of the behavioural LPV framework in Section~\ref{sec:preliminaries}. The LPV Fundamental Lemma and supporting core results are given in Section~\ref{sec:mainresults}, while we show in Section~\ref{sec:linkresults} that the special case of the LPV Fundamental Lemma boils down to the results in \cite{VerhoekAbbasTothHaesaert2021}. We give the conclusions and outlooks in Section~\ref{sec:conclusion}.

\noindent\emph{Notation: }
Let $\mb{A}$ and $\mb{B}$ be vector spaces, the notation $\mb{B}^\mb{A}$ indicates the collection of all maps from $\mb{A}$ to $\mb{B}$. Consider the set $\mb{D}\subseteq\mb{A}\times\mb{B}$ with elements $(a,b)$. The projection of $\mb{D}$ onto the elements of $\mb{A}$ is denoted by $\pi_a\mb{D}\subseteq\mb{A}$, i.e, $\pi_a \mb{D}=\{a\in \mathbb{A} \mid (a,b)\in\mathbb{D}\} $. The degree of a polynomial function $f$ is denoted $\mr{deg}(f)$.
$A_{i,\bullet}$ and $A_{\bullet,j}$ denote the $i$\tss{th} row and the $j$\tss{th} column of a matrix $A\in\R^{n\times m}$, respectively. For a DT signal $w:\Z\to\R^{\dnw}$, we denote its value at discrete time-step $k\in\Z$ by $w(k)$. The forward and backward time-shift operators are denoted as $\q$ and $\q^{\unaryminus 1}$, respectively, such that for a signal $w$, $\q w(k)=w(k+1)$ and $\q^{-1}w(k)=w(k-1)$. For a time-interval $[t_1,t_2]\subset\Z$, the sequence of the values of $w$ on that interval is denoted by $w_{[t_1,t_2]}$, such that $w_{[t_1,t_2]}:[t_1,t_2] \to\R^{\dnw}$. For two trajectories $w_1\in(\mb{R}^{\dnw})^{[t_1,t_2]}$ and $w_2\in(\mb{R}^{\dnw})^{[t_3,t_4]}$, the concatenation of $w_1$ and $w_2$, such that $t_3=t_2+1$, is denoted $w_1\land w_2$. The Hankel matrix of the data-sequence $\tilde{w}:=w_{[1,T]}$, with $t_1$ block rows is denoted by
\begin{equation*}
\mc{H}_{t_1,t_2}(\tilde{w}):= \!\begin{bmatrix}
	\tilde{w}(1) & \tilde{w}(2) & \!\cdots\! & \tilde{w}(t_2) \\
    \tilde{w}(2) & \tilde{w}(3) & \!\cdots\! & \tilde{w}(t_2+1) \\
    \vdots & \vdots & \!\ddots\! & \vdots \\
    \tilde{w}(t_1) & \tilde{w}(t_1+1) & \!\cdots\! & \tilde{w}(t_1+t_2-1)
\end{bmatrix},
\end{equation*}
where $t_2\le T-t_1+1$ and $t_1,t_2>0$. We denote with $\mc{H}_{t_1}(\tilde{w})$, the Hankel matrix  $\mc{H}_{t_1,t_2}(\tilde{w})$ with the maximal possible number of columns, i.e., $t_2= T-t_1+1$. The vector of the values of $w_{[1,\,T]}$ at every time-step is denoted $\mr{vec}(w_{[1,\,T]})=\begin{bmatrix} w^\top\!(1) & \cdots & w^\top\!(T) \end{bmatrix}^\top$. 
\section{Problem statement}\label{sec:problemstatement}
\noindent First, we define a parameter-varying (PV) dynamic system,
\begin{definition}[{PV dynamic system \cite{Toth11_LPVBehav}}]
A PV dynamic system $\Sigma$ is a quadruple $\Sigma = (\mb{T},\mb{P},\mb{W},\mf{B})$
with $\mb{T}$ the time-axis, $\mb{P}\subseteq\mb{R}^{\dnp}$ the scheduling space, $\mb{W}\subseteq\mb{R}^{\dnw}$ the signal space, and $\mf{B}\!\subseteq\!(\mb{W}\!\times\!\mb{P})^\mb{T}$ is the behaviour.
\end{definition}
\noindent In this paper, we consider DT systems, i.e, $\mb{T}=\Z$. Due to linearity of the considered system class with scheduling dependent parameter variations, $\mf{B}$ is linear in the sense that for any $(w,p),(\tilde{w},p)\in\mf{B}$, and $\alpha, \tilde{\alpha}\in\mb{R}$, $(\alpha w+\tilde{\alpha}\tilde{w},p)\in\mf{B}$. Furthermore, $\mf{B}$ is shift invariant, i.e., $\q\mf{B}=\mf{B}$. If $\Sigma$ is not autonomous, we can partition the signal $w$ into a maximal free signal $u$, called the input, with corresponding input space $\mb{U}$, and the signal $y$, called the output, with corresponding output space $\mb{Y}$, satisfying $w=\col(u,y)\in(\mb{U}\times\mb{Y})=\mb{W}$. Note that $y$ does not contain free components, i.e., given $u$, none of the components of $y$ can be chosen freely for every $p\in\pi_p\mf{B}$ \cite{PoldermanWillems1997, Toth2010}. Moreover, in this paper we also consider finite-time trajectories on the time-interval $[t_1,t_2]\subset\mb{Z}$, for which we use the notation 
\begin{multline}\label{eq:behaviourfinite}
\Bfint{}{[t_1,t_2]}:= \big\{(w,p)\in(\mb{W}\times\mb{P})^{[t_1,t_2]}\,\big|\,\exists \,(\omega,\rho)\in\mf{B} \text{ s.t. } \\ (w(t),p(t))=(\omega(t),\rho(t)) \text{ for } t_1\le t\le t_2\big\}.
\end{multline}
\emph{Problem statement:} Given a data-sequence of an unknown LPV system $\Sigma$ with behaviour $\mf{B}$, IO partition $w=\col(u,y)$ and scheduling signal $p$. Under which conditions does the data-sequence span the solution set of the underlying LPV system?


The solution to this problem allows to use a single sequence of data as a data-driven LPV representation in prediction and simulation problems to determine the future response in time.

\section{LPV Behaviours and Representations}\label{sec:preliminaries}
\noindent 
In order to formulate our results we need a brief overview of the LPV behavioural framework \cite{Toth11_LPVBehav, Toth2010} and the introduction of the associated algebraic tools and key representation forms.  
\subsection{Algebraic structure for LPV representations} 
\noindent Let $\mathbb{P}$ be an open subset of $\mathbb{R}^{n_\mathrm{p}}$ and let $\mathcal{R}_\tau(\mathbb{P})$ denote the \emph{set} of real-meromorphic functions of the form $r:\mathbb{P}^\tau \rightarrow \mathbb{R}$ in $n_\mathrm{p} \tau$ variables. For  $\hat{\tau}> \tau$, any  $r \in \mathcal{R}_\tau(\mathbb{P})$ is called equivalent with a $\hat{r} \in \mathcal{R}_{\hat{\tau}}(\mathbb{P})$ if $\hat{r}(\eta_1,\ldots,\eta_{\hat{\tau}})=r(\eta_1,\ldots,\eta_\tau)$ for all $\eta_1,\ldots,\eta_{\tau}\in\mathbb{P}$, as $\hat{r}$ is not  \emph{essentially dependent} on its arguments. Define the set operator $\circleddash$, such that $\mathcal{R}_{\tau+1}(\mathbb{P}) \circleddash  \mathcal{R}_{\tau}(\mathbb{P}) $ contains all $r\in \mathcal{R}_{\tau+1}(\mathbb{P}) $ not equivalent with any element of   $\mathcal{R}_{\tau}(\mathbb{P})$. This prompts to considering the set $\mathcal{R}(\mathbb{P})= \bigcup_{\tau=0}^{\infty} \mathcal{R}_\tau(\mathbb{P}) \circleddash  \mathcal{R}_{\tau-1}(\mathbb{P})$ where $\mathcal{R}_{0}(\mathbb{P})=\mathbb{R}$ and $ \mathcal{R}_{-1}(\mathbb{P})=\emptyset$. We can define addition and multiplication in $\mathcal{R}(\mathbb{P})$ analogous to that of \cite{Toth11_LPVBehav}: if $r_1, r_2 \in \mathcal{R}(\mathbb{P})$, then $r_i \in \mathcal{R}_{\tau_i} (\mathbb{P}) \circleddash   \mathcal{R}_{\tau_i-1}(\mathbb{P})$, for some integer $\tau_i \geq 0$, $i=1,2$, and, by taking $\tau=\max\{\tau_1,\tau_2\}$, the equivalence described above implies that there exist equivalent representations of these functions in $\mathcal{R}_\tau(\mathbb{P})$. Then $r_1+r_2$, $r_1 \cdot r_2$ can be defined as the usual addition and multiplication of functions in $ \mathcal{R}_\tau(\mathbb{P})$ and the result, in terms of the equivalence, is considered to be a $r\in  \mathcal{R}(\mathbb{P})$. For a $p \in \mathbb{P}^\mathbb{Z}$ and $r \in \mathcal{R}(\mathbb{P})$, $r \diamond p: \mb{Z} \rightarrow \mathbb{R}$ is 
\begin{equation*}
(r\diamond p)(k) =r\big(p(k),p(k+1),p(k - 1),\dots, p(k - \tfrac{\tau - 1}{2})\big),
\end{equation*}
where $\tau>0$ is an odd integer such that $r \in \mathcal{R}_{\tau}(\mathbb{P})\ \circleddash \ \mathcal{R}_{\tau-1}(\mathbb{P})$. Similar definition can be given if $\tau$ is even with the last argument being  $p(k+\tfrac{\tau}{2})$. It can be shown that $\mathcal{R}(\mathbb{P})$ is a field. We denote by $\mathcal{R}^{n \times m}(\mathbb{P})$ the set of all $n \times m$ matrices whose entries are elements of $\mathcal{R}(\mathbb{P})$ which also extends the operator $\diamond$ to matrices whose entries are functions from $\mathcal{R}(\mathbb{P})$. It is an important property that multiplication of $\diamond$ with $\q$ is not commutative, in other words, $\q(r\diamond p) \neq (r\diamond p)\q$. To handle this multiplication, for $r\in\mathcal{R}(\mathbb{P})$ we define the shift operations $\overrightarrow{r}, \overleftarrow{r}$ such that $\q(r\diamond p)=(\overrightarrow{r}\diamond p)\q$, $\q^{-1}(r\diamond p)=(\overleftarrow{r}\diamond p)\q^{-1}$ where $\overrightarrow{r}, \overleftarrow{r} \in \mathcal{R}(\mathbb{P})$ s.t. $(\overrightarrow{r}\diamond p)(t)=(r\diamond p)(t+1)$ and $(\shiftl{r}\diamond p)(t)=(r\diamond p)(t-1)$.

Next, we define the algebraic structure of the representations that we use to describe LPV systems, which allows us to use the associated operations to prove our main result. Introduce $\mathcal{R}[\xi]$ as all polynomials in the indeterminate $\xi$ with coefficients in $\mathcal{R}(\mathbb{P})$. $\mathcal{R}[\xi]$ is a ring as it is a general property of polynomial spaces over a field, that they define a ring. With the above defined non-commutative multiplicative rules, $\mathcal{R}[\xi]$ defines an Ore algebra and it is a left and right Euclidean domain \cite{Toth11_LPVBehav}.
Finally, let $\mathcal{R}[\xi]^{n\times m}$ denote the set of matrix polynomial functions with elements in $\mathcal{R}[\xi]$. 
\subsection{Kernel representations} 
\noindent Using $\mathcal{R}[\xi]$ and the operator $\diamond$, we are now able to define a PV difference equation or so-called kernel representation: 
\begin{definition}[PV difference equation \cite{Toth2010}] Consider $R(\xi)=\sum_{i=0}^{n}\! r_i\xi^i\in\mathcal{R}[\xi]^{n_\mathrm{r}\times n_\mathrm{w}}$ and $(w,p)\in(\mathbb{R}^{n_\mathrm{w}}\times\mathbb{R}^{n_\mathrm{p}})^\mathbb{Z}$.
\begin{equation} \label{eq:ch3:01}
(R(\q)\diamond p)w:={\textstyle\sum_{i=0}^{n}}(r_i\diamond p)\q^{i}w=0
\end{equation}
is a PV difference equation with order $n=\mathrm{order}(R)$.
\end{definition}
\noindent The associated behaviour is defined as follows. 
\begin{definition}[{KR-LPV representation \cite{Toth2010}}]\label{def:LPVKERREP}
	The PV difference equation \eqref{eq:ch3:01} is a kernel representation, denoted by $\KR$, of the LPV system $\Sigma = (\Z,\mb{P}\subseteq\R^{\dnp},\R^{\dnw}, \mf{B})$ with scheduling variable $p$ and signals $w$, if
	\begin{equation}\label{eq:lpvkerrep}
		\mf{B}=\setdefinition{(w,p)\in(\R^{\dnw},\mb{P})^\Z\vphantom{_\Z}}{\mdp{R(\q)}w=0},
	\end{equation}
	where $R\in\mc{R}[\xi]^{\cdot\times\dnw}$.
\end{definition}
\noindent From \cite[Thm. 3.6]{Toth2010} we know that for any kernel $R$ in \eqref{eq:lpvkerrep}, there always exists a $\KR$ with full row rank. The order of the kernel representation is the degree of $R$, i.e., $n$ in \eqref{eq:ch3:01}.
The set of admissible scheduling trajectories is denoted by $\mf{B}_\mb{P}=\pi_p\mf{B}$.
The projected behaviour that defines all the signal trajectories compatible with a given fixed scheduling trajectory $p\in\mf{B}_\mb{P}$ is denoted $\mf{B}_p=\setdefinition{w\in\mb{W}^\mb{Z}}{(w,p)\in\mf{B}}$.
Finite time intervals for these sets are denoted as in \eqref{eq:behaviourfinite}.
\subsection{Input-output and state-space representations} 
\noindent 
The behaviours associated with the following representations are required for our main result.
\begin{definition}[{LPV-IO representation \cite{Toth2010}}]\label{def:LPVIOREP}
The IO representation of $\Sigma = (\Z,\mb{P}\subseteq\R^{\dnp},\R^{\dnu+\dny}, \mf{B})$ with IO partition $w=\col(u,y)$ and scheduling $p$ is denoted by $\IOR$ and defined as a parameter-varying difference-equation with order $n_\mr{a}$, where for any $(\col(u,y),p)\in\mf{B}$,
\begin{equation}\label{eq:lpviorep}
\textstyle{\sum_{i=0}^{n_\mr{a}}}\mdp{a_i}\q^iy={\textstyle\sum_{j=0}^{n_\mr{b}}}\mdp{b_j}\q^ju,
\end{equation}
with $a_i\in\mc{R}^{\dny\times\dny}$ and $b_j\in\mc{R}^{\dny\times\dnu}$, $a_{n_\mr{a}}\neq0$ and $b_{n_\mr{b}}\neq0$ being the meromorphic parameter-varying coefficients of the matrix polynomials $R_\mr{u}(\xi)=\sum_{j=0}^{n_\mr{b}}b_j\xi^j$ and full rank $R_\mr{y}(\xi)=\sum_{i=0}^{n_\mr{a}}a_i\xi^i$ with $n_\mr{a}\ge n_\mr{b}\ge0$ and $n_\mr{a}>0$.
\end{definition}
\noindent Finally, we introduce the LPV-SS representation.
\begin{definition}[{LPV-SS representation \cite{Toth2010}}]\label{def:LPVSSREP}
The SS representation of $\Sigma=(\Z,\mb{P}\subseteq\R^{\dnp},\R^{\dnu+\dny}, \mf{B})$ is denoted by $\SSR$ and defined by a first-order PV difference equation in the latent (i.e., state) variable $x:\Z\to\X\subseteq\R^{\dnx}$, with $\X$ the state-space,
\begin{equation}\label{eq:lpvss}
\q x = \mdp{A} x + \mdp{B}u; \ \ y = \mdp{C} x + \mdp{D}u,
\end{equation}
where $(u,y)$ is the IO partition of $\Sigma$, the manifest behaviour
\begin{equation}\label{eq:ssbehaviour}
\mf{B}_{\mr{SS}}\!=\!\{(\col(u,y),p)\in\mf{B}\mid\exists\, x\in (\mb{X})^\Z \text{ s.t. \eqref{eq:lpvss} holds}\},\hspace{-1.2mm}
\end{equation}
is such that $\mf{B}=\pi_{(u,p,y)}\mf{B}_{\mr{SS}}$. Moreover, $A\in\mc{R}^{\dnx\times\dnx}$, $B\in\mc{R}^{\dnx\times \dnu}$, $C\in\mc{R}^{\dny\times\dnx}$, and $D\in\mc{R}^{\dny\times \dnu}$.
\end{definition}
\noindent Next, some integer invariants of the behaviours associated with the representations are introduced. Let $\nBf$ denote the \emph{minimal state dimension} among all $\SSR$ qualifying as a representation of $\mf{B}$. As in \cite{WillemsRapisardaMarkovskyMoor2005}, the \emph{lag} is denoted by $\LBf$, and is the smallest possible lag over all kernel representations $\KR$, i.e., $\LBf$ is equal to the order of a minimal $\KR$. The lag for $\IOR$ is equal to the order $n_\mr{a}$ in Definition \ref{def:LPVIOREP}. Furthermore, note that $\nBf\ge\LBf$ in the MIMO case, while in the SISO case $\nBf=\LBf$. 
\subsection{Notions of minimality, observability and reachability} 
\noindent For $\SSR$, we introduce the notions of observability and reachability in the \emph{almost everywhere} sense, i.e., structural state-observability\fshyp{}reachability\footnote{\emph{Complete} state-observability\fshyp{}reachability is defined in the everywhere sense and is a stronger property than structural state-observability\fshyp{}reachability, and we have complete state-observability\fshyp{}reachability implies structural state-observability\fshyp{}reachability. However, structural state-observability\fshyp{}reachability is a necessary and sufficient property to generate the respective canonical forms \cite{Toth2010}.}, followed by the concepts of minimality for the aforementioned representations. We start with the notion of structural observability, for which we need the $n$-step state-observability matrix function:
\begin{definition}[{Observability matrix \cite{Toth2010}}]
\label{def:nstepobserv}
The $n$-step state-observability matrix $\ms{O}_{n}\in\mc{R}^{ n \dny \times \dnx}$ of $\SSR$ with state dimension $\dnx$ is defined as $\ms{O}_{n} = \begin{bmatrix} \msf{o}_1^\top \!\! &\!\! \msf{o}_2^\top \!\!&\!\! {\scriptstyle\cdots} \!\!&\!\! \msf{o}_n^\top \end{bmatrix}^\top$, with $\msf{o}_1 \!=\! C\!\in\mc{R}^{\dny\times\dnx}$ and $\msf{o}_{i+1}\!=\!\overrightarrow{\msf{o}_i}A\!\in\mc{R}^{\dny\times\dnx}$ for all $i\!>\!1$.
\end{definition}
\noindent With the $n$-step state-observability matrix function, we can define structural observability as follows,
\begin{definition}[Structural state-observability \cite{Toth2010}]
$\SSR$ with state dimension $\dnx$ is called structurally state-observable if its $\dnx$-step observability matrix $\ms{O}_{\dnx}\!$ is full (column) rank.
\end{definition}
\noindent This is full rank in the functional sense as it does not guarantee that $\ms{O}_{\dnx}$ is invertible for all $t\in\mb{Z}$ and $p\in\mf{B}_\mb{P}$. Note that for $\SSR$, $\LBf$ is the minimum integer for which $\rankdef{\ms{O}_{\LBf}}=\dnx$ over all $p$ in an almost everywhere sense. Therefore, let $\Pssobs{L}\subseteq\mb{P}^{\Z}$, associated with a structurally state observable $\SSR$, denote the set of scheduling sequences for which $\rankdef{\!\mdp{\ms{O}_{\LBf}}\!(k)\!}=\dnx$ for all $k\in\Z$, i.e., 
\begin{multline}\label{eq:Pssobs}
\hspace{-3mm}\Pssobs{L}:= \left\{\vphantom{\mc{R}_{\LBf}^{n\dny\times\dnx}} p\in\mb{P}^{\Z}\,\right|\,\rankdef{\!\!\vphantom{O_{O_p}}\mdp{\ms{O}_{\LBf}}\!(k)\!}=\dnx\text{ for }  \\
\hspace{-3mm}\left. L\ge\LBf, \text{and } \forall k\in\Z\vphantom{p\in\mb{P}^{[1,\dnl]}}\text{, with } \ms{O}_{\LBf}\!\in\mc{R}_{\LBf}^{n\dny\times\dnx}(\mb{P})\right\}\!.\hspace{-2.5mm}
\end{multline}
Note that for $L\le\LBf,\,\Pssobs{L}=\{0\}$. Also, for an appropriate measure $\mu$ on $\mb{P}^\Z$, $\mu(\mb{P}^\Z\setminus\mc{P}_{\mr{SS}}^{(\mr{obs})})=0$, when $L\ge\LBf$, corresponding to the almost everywhere sense of structural observability. Structural reachability can be defined in a similar fashion. We first define the $n$-step state-reachability matrix function.
\begin{definition}[{Reachability matrix \cite{Toth2010}}] \label{def:nstepreach}
The $n$-step state-reachability matrix function $\ms{R}_{n}\in\mc{R}^{\dnx\times n \dnu}$ of $\SSR$ with state dimension $\dnx$ is defined as $ \ms{R}_{n} = [ \msf{r}_1 \ \msf{r}_2 \ {\scriptstyle\cdots} \ \msf{r}_n]$, with $\msf{r}_1 \!=\! B\in \mc{R}^{\dnx\times\dnu}$ and $\msf{r}_{i+1}\!=\!A\overleftarrow{\msf{r}_i}\in\mc{R}^{\dnx\times\dnu}$ for all $i\!>\!1$.
\end{definition}
\noindent With the $n$-step state-reachability matrix function, we can define structural reachability as follows,
\begin{definition}[Structural state-reachability \cite{Toth2010}]
$\SSR$ with state dimension $\dnx$ is called structurally state-reachable if its $\dnx$-step reachability matrix $\ms{R}_{\dnx}$ is full (row) rank.
\end{definition}
\noindent This is full rank in the functional sense as it does not guarantee that $\ms{R}_{\dnx}$ is invertible for all $t\in\mb{T}$ and $p\in\mf{B}_\mb{P}$.
We are now ready to define minimality of $\SSR$.
\begin{theorem}[{Induced minimality \cite{Toth2010}}]
The representation $\SSR$ is induced minimal if and only if it is structurally state-observable and it is state-trim, i.e., for all $\mt{x}\in\mb{X}$ there exists a $(u,x,y,p)\in\mf{B}_{\mr{SS}}$ such that $x(0)=\mt{x}$.
\end{theorem}
\noindent This result yields the following definition of minimality for a SS representation $\SSR$.
\begin{definition}[Minimality \cite{Toth2010}]
The $\SSR$ is minimal if the representation is induced minimal and structurally state-reachable. 
\end{definition}
\noindent Minimality in terms of a $\KR$ is that $R\in\mc{R}[\xi]^{\dnr\times\dnw}$ has full row rank, i.e., $\rank(R)=\dnr$. The minimal degree of $\KR$ is the \emph{order} of the system, and is the highest polynomial degree in the rows of $R$ of a minimal $\KR$, i.e., the order is equal to $\LBf$.
 We are now ready to present our main results.

\section{Main results}\label{sec:mainresults}
\noindent First we show results on the continuation of initial trajectories, which will allow the characterisation of the dimensionality of a behaviour.%
%
\subsection{Dimensionality of the restricted behaviour} 
\noindent The $n$\tss{th} impulse response coefficient of an LPV system $\Sigma$ based on its $\SSR$ is 
\begin{equation*}
{\textstyle
h_n = \left\{\begin{array}{ll}
    0, & \text{if }n<0, \\
    D, & \text{if }n =0, \\
    \shiftrn{C}{n}\prod_{i=1}^{n-1}\shiftrn{A}{i}B, & \text{if } n>0.
    \end{array}\right.}
\end{equation*}
The Toeplitz matrix containing the impulse response coefficients of $\Sigma$ is defined as follows
\begin{equation}\label{eq:toeplitzLPV}
\mc{T}_{t_1} := \begin{bmatrix}
    h_0     & 0            & 0             & \cdots & 0 \\
    h_1     & \shiftr{h_0} & 0           & \ddots & \vdots \\
    \vdots  & \vdots       & \!\!\ddots & \!\!\ddots   & 0  \\
    h_{t_1\unaryminus 1} & \shiftr{h}_{\!t_1 \unaryminus 2}     & \cdots   & \shiftrn{h_1}{t_1\unaryminus2}   & \shiftrn{h_0}{t_1\unaryminus1}
\end{bmatrix}.
\end{equation}
If we assume $\SSR$ is \emph{completely} state-observable, there exists always an injective linear map that can be used to reproduce any state, given any $(u,p,y)\in\mf{B}$. However, the notion of \emph{complete} state-observability is rather conservative and the weaker notion of \emph{structural state-observability} is adequate for our purposes. Note that if $\SSR$ is minimal, and thus structurally state-observable, then $\mc{P}_{\mr{SS}}^{(\mr{obs})}$ is trivially non-empty. 
With the following lemma we show that for a finite trajectory there always exists an initial condition when the LPV system admits a SS representation (see e.g. \cite{CoxToth2021} for a similar result).
\begin{lemma}[Initial condition existence]
Let $\SSR$ be a minimal realization of $\Sigma$, with $\mf{B}=\pi_{(w,p)}\mf{B}_{\mr{SS}}$ and IO partitioning $w=\col(u,y)$. For any $(w,p)\in\Bfint{}{[1,T]}$, there exists an $\mt{x}\in\mb{X}$, such that
\begin{equation}\label{eq:existsinit}
\mr{vec}(y) = \mdp{\ms{O}_T}\!(1)\mt{x} + \mdp{\mc{T}_{T}}\!(1)\mr{vec}(u).
\end{equation}
\begin{proof}\pfinapp{%
$\Longleftarrow$: Take any $x(1)=\mt{x}\in\mb{X}$ and any $(u,p)\in\pi_{(u,p)}\Bfint{}{[1,T]}$. As \eqref{eq:lpvss} is a representation of $\mf{B}$, the evolution of the trajectories are governed by \eqref{eq:lpvss}. By definition, \eqref{eq:existsinit} is a recursive application of \eqref{eq:lpvss}, hence $\mr{vec}(y)$ has to satisfy \eqref{eq:existsinit}.\\
$\implies$: As $(w,p)$ is part of the restricted behaviour $\Bfint{}{[1,T]}$, it has a completion in $\mf{B}$. Therefore, for any $(w,p)$, there exists a state trajectory $x\in\pi_x\Bfint{\mr{SS}}{[1,T]}$ associated with $(w,p)$. Taking $\mt{x}=x(1)$ of that state trajectory necessarily satisfies \eqref{eq:existsinit}.%
%
}{See Appendix \ref{appendix:pflemanyinit}.}
\end{proof}
\label{lem:anyinit}
\end{lemma}
\noindent Note that in Lemma \ref{lem:anyinit}, the associated state trajectory $x$, and thus $\mt{x}$, is \emph{not} necessarily unique. 
\begin{lemma}[Initial condition uniqueness]\label{lem:uniqueinit}
Let $\SSR$ be a minimal SS realization of $\Sigma$, such that $\mf{B}=\pi_{(u,p,y)}\mf{B}_{\mr{SS}}$. Then for all $(w_{\mr{ini}}, p_{\mr{ini}})\in\Bfint{}{[1,T_\mr{ini}]}$, where $T_{\mr{ini}}\ge \LBf$ and $p_{\mr{ini}}\in\Pssobs{T_{\mr{ini}}}$, 
\begin{equation}\label{eq:lem2:eq1}
(w_{\mr{ini}}, p_{\mr{ini}}) \land (\col(u_\mr{r},y_\mr{r}),p_\mr{r})\in\Bfint{}{[1,T_\mr{ini}+T_\mr{r}]}
\end{equation}
implies that there is a unique $\mt{x}\in\mb{X}$, such that
\begin{equation}\label{eq:lem2:eq2}
\mr{vec}(y_\mr{r}) = (\ms{O}_{T_\mr{r}} \diamond p_\mr{r})(1)\mt{x} + (\mc{T}_{T_\mr{r}}\diamond p_\mr{r})(1)\mr{vec}(u_\mr{r}).
\end{equation}
\begin{proof}\pfinapp{
Given initial trajectories $w_{\mr{ini}}$ and $p_{\mr{ini}}$, we need to prove the existence of a unique initial vector $\mt{x}\in\mb{X}$, such that the implication holds, for all $u_\mr{r}\in\mb{U}^{[1,T_\mr{r}]}$. We do this constructively. Let $\col(u_{\mr{ini}},y_{\mr{ini}})$ be an IO partitioning of $w_{\mr{ini}}$ and observe that
\begin{multline}
(w_{\mr{ini}}, p_{\mr{ini}}) \land (\col(u_\mr{r},y_\mr{r}),p_\mr{r})\in\Bfint{}{[1,T_\mr{ini}+T_\mr{r}]} \\ \implies (w_{\mr{ini}}, p_{\mr{ini}})\in\Bfint{}{[1,T_\mr{ini}]}
\end{multline}
Since $(w_{\mr{ini}}, p_{\mr{ini}})$ is a trajectory of $\Bfint{}{[1,T_\mr{ini}]}$, it follows from Lemma \ref{lem:anyinit} that there exists some $\bar{\mt{x}}$ such that
\begin{equation}\label{eq:lemuniq:initoutputeq}
\mr{vec}(y_{\mr{ini}}) = (\ms{O}_{T_{\mr{ini}}} \diamond p_{\mr{ini}})(1)\bar{\mt{x}} + (\mc{T}_{T_{\mr{ini}}}\diamond p_{\mr{ini}})(1)\mr{vec}(u_{\mr{ini}}).
\end{equation}
Since $\SSR$ is minimal, $p_{\mr{ini}}\in\Pssobs{T_{\mr{ini}}}$ and $T_{\mr{ini}}\ge \LBf$ imply that the $T_{\mr{ini}}$-step observability matrix $\ms{O}_{T_{\mr{ini}}}$ is full column rank over $p_{\mr{ini}}$. Therefore, \eqref{eq:lemuniq:initoutputeq} has a unique solution in terms of $\bar{\mt{x}}=x(1)$. The initial condition $\mt{x}$ is equal to the state $x(T_{\mr{ini}}\! +\! 1)$, i.e.
\begin{multline}
\mt{x} = x(T_{\mr{ini}} + 1) = \left({\textstyle\prod_{k=0}^{T_{\mr{ini}}-1}}(A\diamond p_{\mr{ini}})(T_{\mr{ini}}-k)\right)x(1)+ \\
+ \begin{bmatrix} \msf{r}_1 & \cdots & \msf{r}_{T_\mr{ini}}
\end{bmatrix}\mr{vec}(u_{\mr{ini}}).
\end{multline}
where $\msf{r}_{T_{\mr{ini}}}\!=\!\mdp{B}(T_{\mr{ini}})$ and $\msf{r}_i\!=\! \mdp{A}(T_{\mr{ini}})\shiftl{r}_{\!\!i+1}$. Uniqueness of $\mt{x}$ follows from uniqueness of $\bar{\mt{x}}$.
}{See Appendix \ref{appendix:pflemuniqueinit}.}
\end{proof}
\end{lemma}
\noindent We can now characterise the dimensionality of $\Bfint{}{[1,L]}$.
\begin{corollary}[Behaviour dimensionality]
Let $\SSR$ be such that $\mf{B}=\pi_{(u,p,y)}\mf{B}_{\mr{SS}}$. Then, $\dim(\Bfint{}{[1,L]})=\dnu L + \nBf$ if and only if $L\ge\LBf$.
\begin{proof}
For any $L\in\mb{N}$, we know from Lemma \ref{lem:anyinit} that there always exists some $\mt{x}$, such that
\begin{equation}\label{eq:col:LgeLbf}
(u,p,\mdp{\ms{O}_L}\mt{x} + \mdp{\mc{T}_{L}}u=y) \in \Bfint{}{[1,L]}.
\end{equation}
Hence, $\dim(y)=\dim(\mdp{\ms{O}_L}\mt{x})+\dim(\mdp{\mc{T}_{L}}u)\ge\dim(\Bfint{}{[1,L]})$ for any $L$. However, when $L\ge\LBf$, there is a \emph{unique} $\mt{x}$ such that \eqref{eq:col:LgeLbf} holds from Lemma \ref{lem:uniqueinit}. Hence, $\dim(y)=\dim(\mdp{\ms{O}_L}\mt{x})+\dim(\mdp{\mc{T}_{L}}u)\le\dim(\Bfint{}{[1,L]})$. Therefore, $\dim(\Bfint{}{[1,L]})=\dnu L + \nBf$.
\end{proof}\label{cor:dimBeh}
\end{corollary}
\subsection{Fundamental Lemma of LPV systems} 
\noindent Consider $\mf{B}_{\mr{SS}}$ associated with a minimal $\SSR$, i.e., $\SSR$ is structurally observable and reachable. We follow the same steps of reasoning as in \cite{WillemsRapisardaMarkovskyMoor2005}. 

\noindent\emph{1) The module of annihilators: }
The module of \emph{annihilators} in the Ore algebra can be seen as the collection of all kernel type of representations of a given $\mf{B}$:
\begin{equation}\label{eq:annihilator1}
\AnnB{}:= \setdefinition{n\in\mc{R}^{\dnw}[\xi]}{n^\top(\q)\odot\mf{B}=0}
\end{equation}
where the notation $n^\top(\q)\odot\mf{B}=0$ means
\begin{align*}
n^\top(\q)\odot\mf{B}=0 \iff \mdp{n(\q)}^\top w=0, \ \foralmostall (w,p)\in\mf{B},
\end{align*}
where $\foralmostall$ indicates for all $(w,p)\in\mf{B}$ in the almost everywhere sense. Similar to \cite{WillemsRapisardaMarkovskyMoor2005}, we require a `special' submodule of the annihilators in \eqref{eq:annihilator1}. Let for $\tau\in\mb{N}$,
the annihilators of degree less than $\tau$ be defined as 
\begin{equation}\label{eq:annihilator2}
\AnnB{\tau}:= \{n\in\mc{R}^{\dnw}[\xi]\mid n\in\AnnB{}, \ \deg(n)\le\tau\}
\end{equation}
Using the notion of the annihilators, we can show the following important property.
\begin{corollary}[Annihilator dimensionality]\label{cor:dimAnn}
Let $\SSR$ be such that $\mf{B}=\pi_{(u,p,y)}\mf{B}_{\mr{SS}}$. If $L\ge\LBf$, then $\dim(\AnnB{L-1})=\dny L - \nBf$.
\begin{proof}
First note that by \cite[Cor. 4.3, Sec. 4.2]{Toth2010}, the SS representation $\SSR$  can always be rewritten into a minimal kernel representation with behaviour $\mf{B}'$ and kernel matrix $R'\in\mc{R}[\xi]^{\dny\times\dnw}$, where $\deg(R')=\dnx$, such that $\mf{B}'=\mf{B}$ in the almost everywhere sense \cite[Thm. 8.7]{Willems1991paradigms}, due to algebraic structure of the behavioural LPV framework. Based on the rows  $[R']_{i, \bullet}$ of $R'$, we can define
its structure indices as $(\mbf{L}_1, \mbf{L}_2, \ldots, \mbf{L}_{\dny})$ with $\mbf{L}_i:=\deg ([R']_{i, \bullet})$. These rows form the basis of the annihilator. More precisely, similarly to \cite[Lem. 4]{MarkovskyDorfler2020} we can generate a matrix $\mathcal{M}$ whose rows span $\AnnB{L-1}$ by populating it with rows $\xi^i [R']_{j, \bullet} $ for $i=0,{\scriptstyle\ldots\,},L-1-\mbf{L}_j$, and $j=1,{\scriptstyle\ldots\,}, \dny$. If $L-\mbf{L}_i\geq 0$ for all $i=1,{\scriptstyle\ldots\,},\dny$, or equivalently if $L\ge\LBf$, then this leads to a full row rank matrix $\mathcal{M}$ with $\dny L- {\textstyle\sum_{i=1}^{\dny}}\mbf{L}_i$ rows.
Hence, $$\dim \AnnB{L-1}=\row \rank (\mathcal{M}) = \dny L- {\textstyle\sum_{i=1}^{\dny}}\mbf{L}_i.$$ Due to the algebraic structure of the LPV behavioural framework, the following result from \cite{Willems1991paradigms} for LTI systems also holds for a minimal $\KR$: $\nBf=\sum_{i=1}^{\dnr}\mbf{L}_i$. Therefore, we have that the $\dny L-\nBf$ linearly independent rows of $\mathcal{M}$ span $\AnnB{L-1}$. Hence, $\dim(\AnnB{L-1})=\dny L - \nBf$.
\end{proof}
\end{corollary}
\noindent\emph{2) Kernel, span and PE: }
We require a more generic notion of the left kernel and the column span of a PV matrix and a generic PE notion. The left kernel of a matrix $M\in\mb{R}^{n\times m}$ 
w.r.t. a $p\in\mf{B}_\mb{P}$ is defined as
\begin{multline}
\mr{Kernel}^{\mr{left}}_{\mc{R}, p}(M) =\left\{ {r\!\in\mc{R}^{1\times n}(\mb{P})}
\right| \\
\left({\textstyle \sum_{i=1}^{n} }r_i M_{i,k} \diamond p \right)(k)=0, \forall k \in[1,m]\bigr\}.
\end{multline}
The column span of $M\!\in\!\mb{R}^{Ln\times m}$ w.r.t. $p\!\in\!\mf{B}_\mb{P}$ is defined as
\begin{multline}
\hspace{-2mm}\mr{Span}^{\mr{col}}_{\mc{R}, p}(M) = \Big\{w\in(\mb{R}^{n})^{[1,L]}\mid\exists\,r_1,\dots,r_m\in\mc{R}^{1\times n}(\mb{P}),\\ 
\text{s.t. }  w_k = {\textstyle \sum_{i=1}^m} (r_i  \bar{M}_{k,i} \diamond p)(k), \forall k \in[1,\,L]\Big\}.
\end{multline}
where $\bar{M}_{k,i}=[M]_{(k-1)n+1:kn,\,i}$.
Observe that 
\begin{equation}\label{eq:anniskerB}
\AnnB{L-1}= {\textstyle\bigcap_{p\in\mathfrak{B}_\mathbb{P}}} \leftpkernel{\Bfint{p}{[1,L]}}\cdot\mf{Q}_{L-1},
\end{equation}
with $\mf{Q}_{L-1}:=[ I \ I\xi \ I\xi^2 \ {\cdots} \ I\xi^{L-1} ]^\top$. From these definitions, we assume the following:
\begin{assumption}[Orthogonality]\label{assumption1}
For a given $M$ and $p$, $\mr{Kernel}^{\mr{left}}_{\mc{R}, p}(M)$ is the orthogonal complement of $\mr{Span}^{\mr{col}}_{\mc{R}, p}(M)$ with respect to $\mc{R}$.
\end{assumption}
\noindent Next, consider the finite trajectories $(w,p)$ of length $T$. The Hankel matrix of depth $L$ associated with $w\in\Bfint{p}{[1,T]}$, i.e., $\mc{H}_L(w)$, has columns that form system trajectories of length $L$, each shifted one time-step. Hence, as $w\in\Bfint{p}{[1,T]}$, any $n\in\AnnB{L-1}$ ensures
\begin{equation}
(\shiftr{n}^{(i)}\diamond p)^{\top}\left[\mc{H}_L(w)\right]_{\bullet,i} = 0,
\end{equation}
for all $i=1, \dots, T-L+1$. The last concept we need to derive the Fundamental Lemma is the notion of PE, which we define w.r.t. a minimal $\KR$ of $\Sigma$ of a given order and dependency class. 
\begin{definition}[PE]\label{def:PE}
The pair $(u,p)\in(\mb{U}\times\mb{P})^{[1,T]}$ is PE of order $L$ w.r.t. to a minimal $\KR$ of order $\le L- 1$ and $\dnr\le\dny$, if for $(\col(u,y),p)\in\Bfint{}{[1,T]}$ it holds that there exists a $[\tau_\mr{s},\tau_\mr{e}]\subseteq[1,T]$, s.t. $\mdp{R(\q)}\!(k)$ is well-defined for all $k\in[\tau_\mr{s},\tau_\mr{e}]$ and for all $\KR$ of a given order, and if there is only one $R$, s.t. for $\tilde{w}=w_{[\tau_\mr{s},\tau_\mr{e}]}$ we have $\mdp{R(\q)}\!(k)\mc{H}_L(\tilde{w})=0$, and $\mc{H}_L(\tilde{w})$ is full row rank.
\end{definition}
\noindent In order to verify the above PE definition in practice, we \emph{need} assumptions on the order and dependency class of the representation of $\Sigma$, see the example in Section~\ref{sec:linkresults} or \cite{DankersTothHeubergerBomboisHof2011} for PE conditions for the specific ARX form.

\noindent\emph{3) The LPV Fundamental Lemma: }
The following result generalises Willems' Fundamental Lemma for LPV systems.
\begin{theorem}[LPV Fundamental Lemma]
Consider the PV system $\Sigma=(\Z,\mb{P}\subseteq\R^{\dnp}, \R^{\dnw}, \mf{B})$ where $\mf{B}=\pi_{(w,p)}\mf{B}_{\mr{SS}}$ for a minimal $\SSR$ with an IO partition ${w}=\col({u},{y})$. Assume Assumption \ref{assumption1} holds and let
$(\tilde{w},\tilde{p})\in\Bfint{}{[1,T]}$ with $\tilde{w}=\col(\tilde{u},\tilde{y})$. If $(\tilde{u},\tilde{p})$ is persistently exciting of order $L+\dnx$ according to Definition \ref{def:PE}, then
\begin{equation}\label{eq:thm:leftkernel}
\leftpkernel[]{\mc{H}_{L}(\tilde{w})}\mf{Q}_{L-1} = \AnnB{L-1},
\end{equation}
where $\mf{Q}_{L-1}:=\begin{bmatrix} I & I\xi & I\xi^2 & \dots & I\xi^{L-1} \end{bmatrix}^\top$, and
\begin{equation}\label{eq:thm:rowspan}
\mr{Span}^{\mr{col}}_{\mc{R}, p}{\mc{H}_{L}(\tilde{w}))} = \Bfint{p}{[1,L]}, \quad \forall p \in \mathfrak{B}_{\mathbb{P}}.
\end{equation}
\begin{proof}
\newcommand{\mfR}{\mf{R}}
\newcommand{\Rel}{r} 
By Assumption \ref{assumption1}, we only have to prove \eqref{eq:thm:leftkernel}. Let 
$$\meu{K}_L := \leftpkernel[]{\mc{H}_{L}(\tilde{w})}\mf{Q}_{L-1}$$ for brevity. The inclusion $\meu{K}_L \supseteq \AnnB{L-1}$ is obvious, as $\mc{H}_{L}(\tilde{w})$ is not guaranteed to fully `contain' $\Bfint{p}{[1,L]}$. Consider the reverse inclusion: $\meu{K}_L\subseteq \AnnB{L-1}$. Assume the contrary, i.e., that there exists some $\Rel$, such that
\begin{equation}
0\neq \Rel = \begin{bmatrix}\Rel_0 & {\scriptstyle \cdots} & \Rel_{L-1} \end{bmatrix} \in \leftpkernel[]{\mc{H}_{L}(\tilde{w})}
\end{equation}
but $\Rel(\xi)=\Rel_0 + \Rel_1\xi+\cdots + \Rel_{L-1}\xi^{L-1}\notin \AnnB{L-1}$. Consider $\mc{H}_{L+\nBf }(\tilde{w})$. Obviously, $\meu{K}_{L+\nBf}$ contains $\AnnB{L+\nBf -1}+\mfR$, with $\mfR\subset \mc{R}^{\dnw}[\xi]$ the (normal) linear span over $\mc{R}$ of
\begin{equation}\label{eq:thm:spanofR}
\mfR = \mr{Span}^{\mr{row}}_\mc{R}\left\{\Rel(\xi), \xi \Rel(\xi), \dots, \xi^{\nBf }\Rel(\xi)\right\}.
\end{equation}
Recall from Corollary \ref{cor:dimAnn} that we have
\begin{equation}
\tdim(\AnnB{L+\nBf -1})=(L+\nBf )\dny-\nBf 
\end{equation}
Clearly, $\tdim(\mfR) = \nBf  + 1$ as \eqref{eq:thm:spanofR} contains $\nBf  + 1$ independent elements by multiplication with $\xi$. We now show that the PE assumption implies $\mfR\cap\AnnB{L+\nBf }\neq \{0\}$. If $\mfR\cap\AnnB{L+\nBf }=\{0\}$, then
\begin{equation}
\tdim(\AnnB{L+\nBf -1}+\mfR)=(L+\nBf)  \dny+1.
\end{equation}
However, the PE condition (Definition \ref{def:PE}) implies that $\tilde{p}\in\Pssobs{L+\nBf}$ and
\begin{multline}
\rank(\mc{H}_{L+\nBf }(\tilde{w}))\ge (L+\nBf )\dnu \ \implies \\ \tdim\Big(\leftpkernel{\mc{H}_{L+\nBf }(\tilde{w})}\Big)\le (L+\nBf )\dny.
\end{multline}
Hence,
\begin{align}
\tdim(\AnnB{L+\nBf -1}+\mfR)& = (L+\nBf )\dny+1 \nonumber \\
&\le \tdim\!\left(\leftpkernel{\mc{H}_{L+\nBf }(\tilde{w})}\right) \nonumber \\
&\le (L+\nBf )\dny. \label{eq:thm:dimAnnplusR}
\end{align}
Therefore $\mfR\cap\AnnB{L+\nBf }\neq \{0\}$. Consequently, there is a linear combination of
\begin{equation}
\Rel^\top(\xi), \xi \Rel^\top(\xi), \dots, \xi^{\nBf }\Rel^\top(\xi),
\end{equation}
that is contained in $\AnnB{L+\nBf }$. In terms of the minimal kernel representation $\mdp{R(\q)}w=0$ of $\mf{B}$, this means that there is a $0\neq f \in \mc{R}[\xi]$, such that $f \Rel = FR$, for some $0 \neq F \in \mc{R}^{1\times \mr{rowdim}(R)}[\xi]$. If $\mr{deg}(f)\ge1$, then there is a $\lambda'\in\mb{C}$ such that $f(\lambda')=0$, hence $F(\lambda')R(\lambda')=0$. Next, we use the fact that $\SSR$ has an equivalent minimal kernel representation based on the Elimination Lemma \cite[Thm. 3.3]{Toth2010}. Furthermore, $\mdp{R(\q)}w=0$ of $\mf{B}$ is a minimal kernel representation, therefore combination of its rows spans $\AnnB{L+\nBf}$. Hence, $f$ can be reduced to $\deg(f)=0$ by cancelling the common factors between $f$ and $F$. Then $\Rel = FR$. This contradicts the assumption $\Rel\notin \AnnB{L\unaryminus1}$. Hence, $\meu{K}_L \subseteq \AnnB{L\unaryminus1}$ and \eqref{eq:thm:leftkernel} holds, concluding the proof.
\end{proof}
\label{thm:lpvfundamentallemma}
\end{theorem}
\begin{remark}
Suppose we obtained the kernel that spans all the annihilators associated with the behaviour, it is possible to construct the (left-)module in $\mc{R}^{\dnr\times\dnw}[\xi]$ generated by the kernel (see \cite[Ch. 4]{Toth2010} for a definition). This module is the building block for all the equivalent minimal SS representations associated with the system. This links our result to subspace identification, see \cite{CoxToth2021} and references therein.
\end{remark}
\section{Fundamental Lemma under affine dependence}\label{sec:linkresults}
\noindent In this section, we discuss Theorem~\ref{thm:lpvfundamentallemma} for the special case of static, affine dependence, which recover the results derived in \cite{VerhoekAbbasTothHaesaert2021}, and give a simulation example for this particular case.
\subsection{Simplified results} 
\noindent Consider an LPV system $\Sigma$ with LPV-IO representation
\begin{equation}\label{eq:sec5:lpvsysverhoek}
y(k)+\sum_{i=1}^{n_\mr{a}}a_i(p(k\unaryminus i))y(k\unaryminus i)=\sum_{i=1}^{n_\mr{b}}b_i(p(k\unaryminus i))u(k\unaryminus i),
\end{equation}
where the functions $a_i, \,b_i$ have affine dependence, i.e., 
\begin{subequations}\label{eq:sec5:defverhoekres}
\begin{align}
\hspace{-1mm}a_i(p(k - i)) &= {\textstyle\sum_{j=0}^{n_\mr{p}}}\,a_{i,j}p_j(k - i), && a_{i,j}\in\mb{R}^{\dny\times\dny}, \hspace{-1mm}\\
\hspace{-1mm}b_i(p(k - i)) &= {\textstyle\sum_{j=0}^{n_\mr{p}}}\,b_{i,j}p_j(k - i), && b_{i,j}\in\mb{R}^{\dny\times\dnu}.\hspace{-1mm}
\end{align}
\end{subequations}
This gives that $\Sigma$ has the behaviour 
\begin{equation*}
    \mf{B}:=\!\setdefinition{(u,p,y)\!\in\!(\R^{\dnu}\!\!\times\!\mb{P}\!\times\!\R^{\dny})^\Z\!}{\text{\eqref{eq:sec5:lpvsysverhoek} holds with \eqref{eq:sec5:defverhoekres}}}.
\end{equation*}
The representation \eqref{eq:sec5:lpvsysverhoek} under the considered affine dependence \eqref{eq:sec5:defverhoekres} can be rewritten as an
implicit LTI form \cite{VerhoekAbbasTothHaesaert2021}
\begin{equation}\label{eq:sec5:lpviorecast}
E\msf{y}(k)+{\textstyle \sum_{i=1}^{n_\mr{a}}}A_i\msf{y}(k\unaryminus i)={\textstyle \sum_{i=1}^{n_\mr{b}}}B_i\msf{u}(k\unaryminus i), 
\end{equation}
with $E=[I \ 0]$, $A_i=\begin{bmatrix} a_{i,0} & \cdots & a_{i,n_\mr{p}}\end{bmatrix}$ similar $B_i$, and
\begin{equation}\label{eq:sec5:auxIO}
\msf{u}(k) := \begin{bsmallmatrix}
u(k) \\
p(k) \kron u(k)
\end{bsmallmatrix}, \quad \msf{y}(k) := \begin{bsmallmatrix}
y(k) \\
p(k) \kron y(k)
\end{bsmallmatrix},
\end{equation}
with $\kron$ the Kronecker product. For this special case, Theorem~\ref{thm:lpvfundamentallemma} and the application of the LTI 
Fundamental Lemma (adapted for \eqref{eq:sec5:lpviorecast} in \cite{VerhoekAbbasTothHaesaert2021}) both give
\begin{equation}\label{eq:sec5:LPV_pred}
\begin{bmatrix}
\mc{H}_L\left( u \right) 		\\
\mc{H}_{L}\left( p\kron u \right) -\bar{\mc{P}}_{n_\mr{u}}\mc{H}_{L}\left( u \right) \\
\mc{H}_L\left( y \right) 		\\
\mc{H}_{L}\left( p\kron y \right) -\bar{\mc{P}}_{n_\mr{y}}\mc{H}_{L}\left( y \right)
\end{bmatrix}g = \begin{bmatrix}
\mr{vec}(\bar{u}) \\
0	 \\
\mr{vec}(\bar{y}) \\
0	 
\end{bmatrix},
\end{equation}
where $\bar{\mc{P}}_{n}$ is a block-diagonal matrix with diagonal blocks $\bar{p}(k)\kron I_{n\times n}$, $(u,p,y)\in\left.\mf{B}'\right|_{[1,T]}$ and $(\bar{u},\bar{p},\bar{y})\in\left.\mf{B}'\right|_{[1,L]}$. 
\subsection{The link with Theorem \ref{thm:lpvfundamentallemma}} 
\noindent We show how the application of the LTI Fundamental Lemma on \eqref{eq:sec5:lpviorecast} derived in \cite{VerhoekAbbasTothHaesaert2021} result in a special case of
Theorem \ref{thm:lpvfundamentallemma}. Note that with the dependency \eqref{eq:sec5:defverhoekres}, there is a minimal kernel representation of \eqref{eq:sec5:lpvsysverhoek}, i.e.,
\begin{equation}\label{eq:sec5:kerrep}
\mdp{R(\q)}\!(k)w(k) = 0, \quad R(\q)= r_0+\textstyle{\sum_{i=1}^{n}}r_i\q^i,
\end{equation}
with $r_i \in\mc{R}(\mb{P})^{\dnr\times\dnw}$ and $\rank(R)=\dnr$. Hence, for any $\tilde{w}\in\Bfint{\tilde{p}}{[1,L]}$, with $L\ge\nBf$,
\begin{equation*}
\big(\bar{r}\diamond\tilde{p}\big)(1)\cdot[\mc{H}_L(\tilde{w})]_{\bullet,1}=0,
\end{equation*}
where $\bar{r}=\begin{bmatrix} \bar{r}_0 & \dots & \bar{r}_{L-1}\end{bmatrix}$ with $\bar{r}(\xi)=\sum_{i=0}^{L-1}\bar{r}_i\xi^i$ and 
\begin{equation*}
\bar{r}(\xi)\in\mr{Span}_\mc{R}^{\mr{row}}\{R(\xi),\, \xi R(\xi), \dots, \xi^{L-n}R(\xi)\}.
\end{equation*}
Introduce the set of affine coefficients with static dependence (as in \eqref{eq:sec5:defverhoekres}) as $\mc{R}_{\mr{aff}}(\mb{P})$, which is a subclass of $\mc{R}_1(\mb{P})$. Let
\begin{equation*}
\mc{R}_{\mr{aff}}[\xi]:= \{R\in\mc{R}[\xi] \mid R(\xi) = {\textstyle\sum_{i=0}^n} r_i\xi^i,  \, \shiftr{r_i}\in\mc{R}_{\mr{aff}}(\mb{P})\}
\end{equation*}
be the collection of kernel representations with coefficients having shifted affine dependence on $p$. Note that if $R$ is defined as in \eqref{eq:sec5:kerrep}, where $\mdp{r_i}(k)= r_{i,0} + \sum_{j=1}^{\dnp}r_{i,j}p_j(k\unaryminus i)$, with $p_j$ the $j$\tss{th} element of the scheduling vector, then 
\begin{equation*}
\bar{r}(\xi)\!\in\!\mr{Span}_{\mb{R}}^{\mr{row}}\{R(\xi),\xi R(\xi), {\scriptstyle\cdots}, \xi^{L\unaryminus n}R(\xi)\}\!\neq\!\{0\}\!\in\!\mc{R}_{\mr{aff}}[\xi],
\end{equation*}
and having also only affine shifted dependence. Furthermore, this restricted span fulfils all the properties of the proof in Theorem \ref{thm:lpvfundamentallemma}. Therefore, due to Assumption \ref{assumption1}, the orthogonal complement w.r.t. $\mc{R}$ of $ \mr{Span}^{\mr{col}}_{\mc{R}, \tilde{p}}\big(\mc{H}_{L}(\tilde{w})\big)$
of a PE sequence $w\in\Bfint{\mb{W}}{[1,T]}$ of order $L+\dnx$ can also be restricted to $\mc{R}_{\mr{aff}}(\mb{P})$, without loss of generality. This means that
\begin{equation*}
(\bar{r}\diamond\tilde{p})(1) = \begin{bsmallmatrix} \bar{r}_{0,0} + {\textstyle \sum_{j=1}^{\dnp}}\bar{r}_{0,j}\tilde{p}_j(1) && \bar{r}_{1,0} + {\textstyle \sum_{j=1}^{\dnp}}\bar{r}_{1,j}\tilde{p}_j(2) && \cdots & \cdots \end{bsmallmatrix}.
\end{equation*}
Hence, $(\bar{r}\diamond\tilde{p})(1)\,\mc{H}_{L,1}(\tilde{w})=0$ implies $\tilde{r}\,\mc{H}_L\!\begin{psmallmatrix} \tilde{w}\\\tilde{p}\kron\tilde{w} \end{psmallmatrix}=0$, with $\tilde{r}\in\R^{\dnr\times(1+\dnp)\dnw}$, containing all $\bar{r}_{i,j}$. Now, we can repeat the whole derivation for $\tilde{r}\,\mc{H}_L\!\begin{psmallmatrix} \tilde{w}\\\tilde{p}\kron\tilde{w} \end{psmallmatrix}=0$, using the orthogonal complement property under $\R$ as a special case of Theorem \ref{thm:lpvfundamentallemma} (retrieving the original result in \cite{WillemsRapisardaMarkovskyMoor2005}). 
\subsection{Numerical example} 
\begin{figure}[t]
\centering

\includegraphics[width=\linewidth]{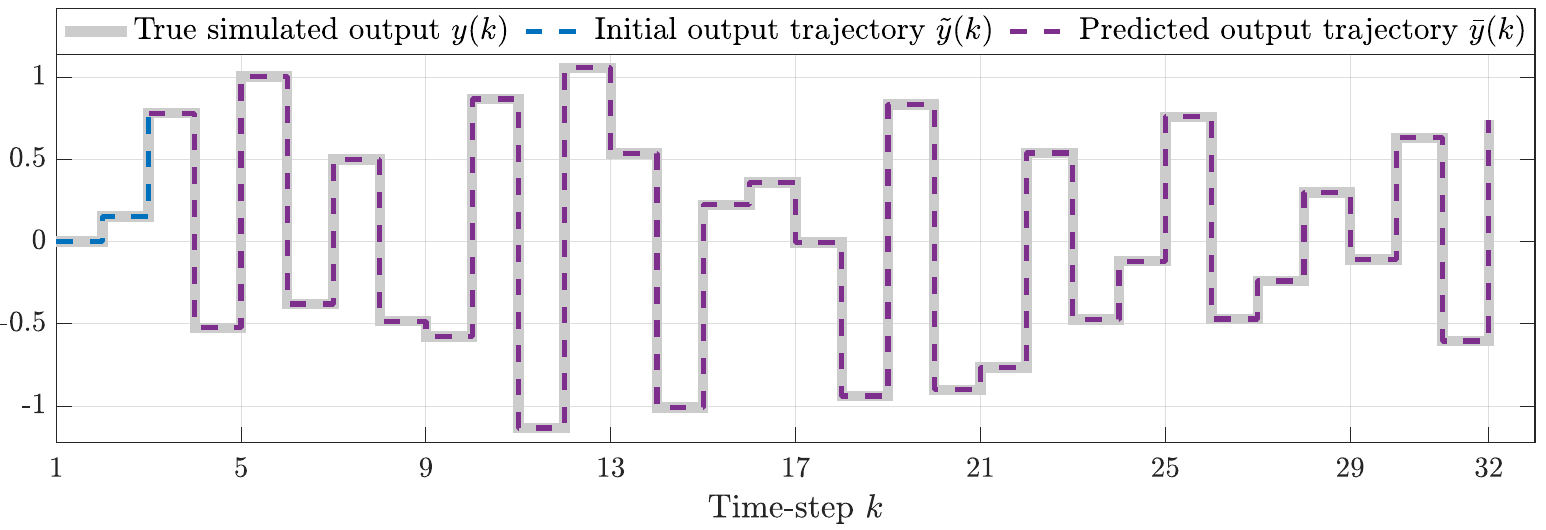}

\vspace{-4mm}

\caption{Results of the simulation problem. The blue coloured data corresponds to the initial trajectory of length $T_{\mr{ini}}$ in Lemma \ref{lem:uniqueinit}. The purple coloured data corresponds to the predicted trajectory of length $L$, which is obtained using Theorem~\ref{thm:lpvfundamentallemma}.}

\vspace{-5mm}

\label{fig:example}
\end{figure}

\noindent We present a simulation example using the SISO LPV system from \cite{VerhoekAbbasTothHaesaert2021} in the form \eqref{eq:sec5:lpvsysverhoek}--\eqref{eq:sec5:defverhoekres} with $n_\mr{a}\!=n_\mr{b}\!=n_\mr{p}\!=2$ and 
\begin{align*}
A_1&=[ 1   \ \unaryminus 0.5 \ \unaryminus 0.1],& \ A_2 &= [ 0.5 \ \unaryminus 0.7 \ \unaryminus 0.1], \\
B_1&=[ 0.5 \ \unaryminus 0.4 \ 0.01           ],& \ B_2 &= [ 0.2 \ \unaryminus 0.3 \ \unaryminus 0.2].
\end{align*}
We use Lemma~\ref{lem:uniqueinit} to simulate the system for $L=30$ steps, given an initial trajectory $(\tilde{u},\tilde{p},\tilde{y})$ of length $T_{\mr{ini}}$, the future input and scheduling trajectories $(\bar{u},\bar{p})$ of length $L$, and a data-dictionary $({u},{p},{y})$ of persistently exciting data. The data-dictionary is generated using a random input and scheduling trajectories of length 193, and is used to represent the `unknown' LPV system using Theorem~\ref{thm:lpvfundamentallemma}. Note that $\LBf=2$, i.e., $T_{\mr{ini}}=2$. We can now solve \eqref{eq:sec5:LPV_pred} for Hankel matrices of depth $T_{\mr{ini}}+L$ in order to obtain the output $\bar{y}$ such that $(\tilde u, \tilde{p},\tilde{y})\land (\bar{u},\bar{p}, \bar y)\in\Bfint{}{[1,T_{\mr{ini}}+L]}$. The results in Fig.~\ref{fig:example}, show that we can reproduce the output exactly for the full horizon $L$, by only solving \eqref{eq:sec5:LPV_pred}, which only contains data-sequences from the unknown LPV system. See \cite{chriswebsite} for more plots and an additional example.
%
%
%
%
%
\section{Conclusions and future work}\label{sec:conclusion}
\noindent By establishing the LPV form of Willems' Fundamental Lemma, we have shown that a single sequence of data generated by an unknown LPV system is sufficient to characterise its behaviour and describe its future responses. We have also shown that in case the system can be represented by an IO representation with simple shifted affine dependency, the Fundamental Lemma results in a simple algebraic relation that can be efficiently used for characterising the future system response. We have illustrated the applicability of the latter relation in a simulation example.
Our result can be seen as a stepping stone towards data-driven analysis and control for general NL systems.
%

\bibliographystyle{IEEEtran}
\bibliography{references}

\pfinapp{}{
\appendix
\subsection{Proof of Lemma \ref{lem:anyinit}}\label{appendix:pflemanyinit}
\begin{proof}
\chris{Note that \eqref{eq:lpvss} is the recursive representation of \eqref{eq:existsinit}.}\\
$\Longleftarrow$: Take any $x(1)=\mt{x}\in\mb{X}$ and any complete $(u,p)\in\pi_{(u,p)}\Bfint{}{[1,T]}$. As \eqref{eq:lpvss} is a representation of $\mf{B}$, the evolution of the trajectories are governed by \eqref{eq:lpvss}. Hence, \chris{via the recursive nature}, the output $\mr{vec}(y)$ needs to satisfy it.\\
$\implies$: When for $(w,p)\in(\mb{W}\times\mb{P})^{[1,T]}$ there exists an $\mt{x}$ such that \eqref{eq:existsinit} is satisfied for $[1,T]$, this implies that $(w,p)$ are governed by \eqref{eq:lpvss} for the interval $[1,T]$. As $\mf{B}$ is fully defined by \eqref{eq:lpvss}, so is $\Bfint{}{[1,T]}$ for the interval $[1,T]$, as there is completion of $(w,p)=\Bfint{}{[1,T]}$ in $\mf{B}$. This implies that $(w,p)$ must be an element of $\Bfint{}{[1,T]}$.
\end{proof}
\subsection{Proof of Lemma \ref{lem:uniqueinit}}\label{appendix:pflemuniqueinit}
\begin{proof}
Given initial trajectories $w_{\mr{ini}}$ and $p_{\mr{ini}}$, we need to prove the existence of a unique initial vector $\mt{x}\in\mb{X}$, such that the implication holds, for all $u_\mr{r}\in\mb{U}^{[1,T_\mr{r}]}$. We do this constructively. Let $\col(u_{\mr{ini}},y_{\mr{ini}})$ be an IO partitioning of $w_{\mr{ini}}$. Note that $w_{\mr{ini}}$ and $p_{\mr{ini}}$ are not free:
\begin{multline}
(w_{\mr{ini}}, p_{\mr{ini}}) \land (\col(u_\mr{r},y_\mr{r}),p_\mr{r})\in\Bfint{}{[1,T_\mr{ini}+T_\mr{r}]} \\ \implies (w_{\mr{ini}}, p_{\mr{ini}})\in\Bfint{}{[1,T_\mr{ini}]}
\end{multline}
Since $(w_{\mr{ini}}, p_{\mr{ini}})$ is a system trajectory of $\Bfint{}{[1,T_\mr{ini}]}$, it follows from Lemma \ref{lem:anyinit} that there exists some $\bar{\mt{x}}$ such that
\begin{equation}\label{eq:lemuniq:initoutputeq}
\mr{vec}(y_{\mr{ini}}) = (\ms{O}_{T_{\mr{ini}}} \diamond p_{\mr{ini}})(1)\bar{\mt{x}} + (\mc{T}_{T_{\mr{ini}}}\diamond p_{\mr{ini}})(1)\mr{vec}(u_{\mr{ini}}).
\end{equation}
Moreover, the assumptions that $\SSR$ is a minimal representation, $p_{\mr{ini}}\in\Pssobs{T_{\mr{ini}}}$ and $T_{\mr{ini}}\ge \LBf$ imply that the $T_{\mr{ini}}$-step observability matrix $\ms{O}_{T_{\mr{ini}}}$ is full column rank over $p_{\mr{ini}}$. Therefore, the system of equations \eqref{eq:lemuniq:initoutputeq} has a unique solution $\bar{\mt{x}}$. The initial condition $\mt{x}$ is equal to the state $x(T_{\mr{ini}}\! +\! 1)$, i.e.
\begin{multline}
\mt{x} = x(T_{\mr{ini}} + 1) = \left({\textstyle\prod_{k=0}^{T_{\mr{ini}}-1}}(A\diamond p_{\mr{ini}})(T_{\mr{ini}}-k)\right)x(1)+ \\
+ \begin{bmatrix} \msf{r}_1 & \cdots & \msf{r}_{T_\mr{ini}}
\end{bmatrix}\mr{vec}(u_{\mr{ini}}).
\end{multline}
where $\msf{r}_{T_{\mr{ini}}}\!=\!\mdp{B}(T_{\mr{ini}})$, $\msf{r}_i\!=\! \mdp{A}(T_{\mr{ini}})\shiftl{r}_{\!\!i+1}$ and $x(1)\!=\bar{\mt{x}}$. Uniqueness of $\mt{x}$ follows from uniqueness of $\bar{\mt{x}}$.
\end{proof}

}

\end{document}